\documentstyle[12pt]{article}
\def\zid{1\kern-0.36em\llap~1}

\newcommand{\beq}{\begin{equation}}
\newcommand{\ber}{\begin{eqnarray}}
\newcommand{\eeq}{\end{equation}}
\newcommand{\eer}{\end{eqnarray}}

\begin{document}

\begin{titlepage}
\rightline{[SUNY BING 7/2/00] } \rightline{ hep-ph/0007086}
\vspace{2mm}
\begin{center}
{\bf ON MEASUREMENT OF HELICITY PARAMETERS IN TOP QUARK DECAY}\\
\vspace{2mm} Charles A. Nelson\footnote{Electronic address:
cnelson @ binghamton.edu  } and L. J. Adler, Jr.\\ {\it Department
of Physics, State University of New York at Binghamton\\
Binghamton, N.Y. 13902-6016}\\[2mm]
\end{center}


\begin{abstract}
To enable an evaluation of future measurements of the helicity
parameters for \newline $t \rightarrow W^+ b$ decay in regard to
$\tilde{T}_{FS}$ violation, this paper considers the effects of an
additional pure-imaginary coupling, $ i g_i / 2 \Lambda_i $ or $ i
g_i $, associated with a specific, single additional Lorentz
structure, $ i = S, P, S \pm P , \ldots $. Sizable $
\tilde{T}_{FS}$ violation signatures can occur for low-effective
mass scales ( $ < 320 GeV $ ), but in most cases can be more
simply excluded by $ 10\% $ precision measurement of the
probabilities $ P(W_L)$ and $P(b_L)$. Signatures for excluding the
presence of $\tilde{T}_{FS}$ violation associated with the two
dynamical phase-type ambiguities are investigated.

\end{abstract}

\end{titlepage}

\section{ Introduction }

In  $t \rightarrow W^+ b $ decay, it is important to be able to
evaluate future measurements of competing observables consistent
with the standard model (SM) prediction of only a $g_{V-A}$
coupling and of only its associated discrete-symmetry violations.
For this purpose, without consideration of possible explicit $
\tilde{T}_{FS}$ violation, in [1] plots were given of the values
of the helicity parameters in terms of a ``$(V-A)$ $+$ Single
Additional Lorentz Structure" versus effective-mass scales for new
physics, $\Lambda_i$, associated with each additional Lorentz
structure.  In this paper, the effects of possible explicit $
\tilde{T}_{FS}$ violation are reported.  In the present
formulation, by ``explicit $ \tilde{T}_{FS}$ violation", c.f.
Sec.2, we mean an additional complex-coupling, $ g_i / 2 \Lambda_i
$ or $ g_i $, associated with a specific single additional Lorentz
structure, $ i = S, P, S \pm P , \ldots $.

The main motivation for the present analysis are the observed $CP$
and $T$ violations in $K^{0}$ decay. Although these
discrete-symmetry violations are empirically well-described by the
CKM matrix which describes the linear superposition of the quark
mass eigenstates which appears in the phenomenological weak
eigenstates, the fundamental origin of these symmetry violations
is still unknown. Experimental results should soon be available
about whether the CKM formulation is also successful in b-quark
decay.  In the case of the strong interactions, there is the
opposite difficulty of a fundamental strong $CP$ problem which has
led to the prediction of the existence of axions, the
Nambu-Goldstone bosons associated with a global $U(1)_{PQ} $
symmetry.  These axions have yet to be discovered.  Lastly, and
perhaps more significantly for t-quark decay, most astrophysics
studies of electroweak baryogenesis conclude that additional
sources of $CP$ violation, beyond CKM, in elementary particle
physics are necessary to explain the observed baryon-to-photon
ratio.  So in spite of the robustness of the standard model and of
the CKM formulation, perhaps after all, t-quark decay is not the
wrong place to look for $CP$ and $T$ violations.

A first measurement of the longitudinal W boson fraction was
reported in [2].  A recent working group review of t-quark physics
is in [3]. A recent review of $CP$ violation in t-quark physics is
in [4]. Besides these references and those listed in [1], some of
the related recent literature is [5-14].

The present analysis assumes that future measurements of $t
\rightarrow W^+ b $ decay will be at least approximately
consistent with the SM prediction of only a $g_{V-A} $ coupling.
If the SM is correct, one expects that the $A(0, -1/2)$ and
$A(-1,-1/2)$ moduli and relative phase $\beta_L$ will be the first
quantities to be somewhat precisely determined. As shown by Table
1, the $\lambda_b = 1/2$ moduli are factors of 30 and 100 smaller
in the SM.  The helicity parameters appear directly in various
polarization and spin-correlation functions for $t \rightarrow W^+
b $ decay such as those obtained in [15]. By measurement of
independent helicity parameters, or from other empirical analyses
of spin-correlation and polarization observables, it will be
possible to test in several independent ways that the R-handed
b-quark amplitudes, $\lambda_b = 1/2$, are indeed negligible to
good precision. Eventually there should also be direct evidence
for their existence if the SM is correct.

If the R-handed amplitudes are negligible, then besides $P(b_L)
\simeq 1$ it follows that \newline $\zeta \simeq
2P(W_{L})-P(b_{L})$ and that $\omega \simeq \eta $. Showing an
approximate empirical absence of R-handed amplitudes would also be
useful in regard to tests for $ \tilde{T}_{FS}$ violation:
Assuming that the L-handed amplitudes dominate, the $\eta
_{L}^{^{\prime }}$ helicity parameter satisfies the relation
\newline
$($ $\eta _{L}^{^{\prime }})^{2}\cong \frac{1}{4}[P(b_{L})+\zeta ][%
P(b_{L})-\zeta ]-($ $\eta _{L})^{2}$.  For instance, in the SM the
vanishing of the right-hand-side is due to the vanishing of $\sin
\beta _{L}$ provided that the R-handed amplitudes are negligible.
If $ \tilde{T}_{FS}$ violation were to occur, besides normally a
non-zero approximate-equality in the above relation, there would
normally also be a non-zero $\omega^{\prime } \simeq \eta^{\prime
} $ if the R-handed amplitudes are negligible.

{\bf Remarks on the dynamical phase-type ambiguities:}

Due the dominance of the L-handed amplitudes in the SM, the
occurrence of the two dynamical ambiguities [1] displayed in lower
part of Table 1 is not surprising because these three chiral
combinations only contribute to the L-handed b-quark amplitudes in
the $m_b \rightarrow 0$ limit. Since pairwise the couplings are
tensorially independent, the $g_{V-A} + g_{S+P} $ and $g_{V-A} +
g_{f_M + f_E} $ mixtures can each be tuned by adjusting a purely
real $\Lambda_i$ to reproduce, with opposite sign, the SM ratio of
the two  $( \lambda_W =0, -1 )$ L-handed amplitudes . Likewise, if
experimental data were to suggest that the R-handed amplitudes are
larger than expected, e.g. $ P(b_L) \neq 1$, this might be due to
the presence of additional $V+A, S-P, f_M - f_E$ type couplings.
Since the $S \pm P$ couplings only contribute to the longitudinal
helicity W amplitudes, they might be of interest in the case of an
unexpected W longitudinal/transverse polarization ratio. Versus
the upper part of Table 1, given the small $m_b$ mass, this is the
reason that the sign of the $A(0,-1/2)$ amplitude can be switched,
without other important changes, by the addition of the $S+P$
coupling.

However, in the case of the $f_M+f_E$ phase-type ambiguity, from
Table 1 there are 3 numerical puzzles versus the SM values.  In
the upper part, the $A_{+} (0,-1/2)$ amplitude for $g_L
+g_{f_M+f_E}$ has about the same value in $g_L = 1 $ units, as the
$A_{SM} (-1,-1/2)$ amplitude in the SM.  As $m_b \rightarrow 0$, $
\frac{A_{+} (-1,-1/2) } { A_{SM} (0,-1/2) } \rightarrow
\frac{m_{t}(m_{t}^{2}-m_{W}^{2})}{\sqrt{2}m_{W}(m_{t}^{2}+m_{W}^{2})}
= 1.0038 $. The other numerical puzzle(s) is the occurrence in the
lower part of the Table 1 of the same magnitude of the two
R-handed b-quark amplitudes $A_{New} = A_{g_L =1} / \sqrt \Gamma $
for the SM and for the case of $g_L +g_{f_M+f_E}$. Except for the
differing partial width, by tuning the magnitude of L-handed
amplitude ratio to that of the SM, the R-handed amplitude's moduli
also become about those of the SM. With $\Lambda_{f_M+f_E}$
determined as in Sec. 3, for the $A_{New}$ amplitudes $ \vert
A_{+} - A_{SM} \vert \sim (m_b / m_t ) ^2 $ versus for instance $
\vert A_{SM} (\lambda_W, 1/2) \vert \propto m_b $. Of course, the
row with SM values is from a ``theory" whereas the row of $g_L
+g_{f_M+f_E}$ values is not. Nevertheless, dynamical SSB and
compositeness/condensate considerations do continue to stimulate
interest [15] in additional tensorial $f_{M}+f_{E}$ couplings. In
Table 1, due to the additional $f_M + f_E$ coupling, the net
result is that it is the $\mu =\lambda _{W^{+} } -\lambda _b = -
1/2 $ helicity amplitudes $A_{New}$ which get an overall sign
change.

Fortunately, a sufficiently precise measurement of the sign of $
\vert \eta_L \vert = 0.46$(SM) due to the large interference
between the W longitudinal/transverse amplitudes can resolve the
$V-A$ and $f_M + f_E$ lines of this table.  Similarly,
sufficiently precise measurements of both  $ \eta_L $ and
${\eta_L}^{'} $ could resolve the analogous dynamical ambiguity in
the case of a partially-hidden  $ \tilde{T}_{FS}$ violation
associated with the additional $f_M + f_E $ coupling, see Figs.8
in Sec. 3. A precise measurement of the partial width $\Gamma$
could also be useful.

\section{Consequences of Explicit $ \tilde{T}_{FS}$ Violation}

For \hskip1em $t \rightarrow W^+ b$, the most general Lorentz
coupling is $ W_\mu ^{*} J_{\bar b t}^\mu = W_\mu ^{*}\bar
u_{b}\left( p\right) \Gamma ^\mu u_t \left( k\right) $ where $k_t
=q_W +p_b $, and
\begin{eqnarray}
\Gamma _V^\mu =g_V\gamma ^\mu + \frac{f_M}{2\Lambda }\iota \sigma
^{\mu \nu }(k-p)_\nu + \frac{g_{S^{-}}}{2\Lambda }(k-p)^\mu
\nonumber \\ +\frac{g_S}{2\Lambda }(k+p)^\mu
+%
\frac{g_{T^{+}}}{2\Lambda }\iota \sigma ^{\mu \nu }(k+p)_\nu
\end{eqnarray}
\begin{eqnarray}
\Gamma _A^\mu =g_A\gamma ^\mu \gamma _5+ \frac{f_E}{2\Lambda
}\iota \sigma ^{\mu \nu }(k-p)_\nu \gamma _5 +
\frac{g_{P^{-}}}{2\Lambda }(k-p)^\mu \gamma _5  \nonumber \\
+\frac{g_P}{2\Lambda }%
(k+p)^\mu \gamma _5  +\frac{g_{T_5^{+}}}{2\Lambda }\iota \sigma
^{\mu \nu }(k+p)_\nu \gamma _5
\end{eqnarray}
In this paper, in consideration of the additional Lorentz
structures to pure $V-A$, we consider the $g_i$ or $\Lambda_i$ as
complex phenomenological parameters. For $g_L = 1$ units with $g_i
= 1$, the nominal size of $\Lambda_i$ is $\frac{m_t}{2} = 88GeV$,
see [1].  In the SM, the EW energy-scale is set from the
Higgs-field vacuum-expectation-value by the parameter
$v=\sqrt{-\mu^2 / \vert \lambda \vert} = \sqrt{2} \langle 0|\phi
|0\rangle \sim 246GeV$.

The helicity formalism is based on the assumption of Lorentz
invariance but not on any specific discrete symmetry property of
the fundamental amplitudes, or couplings.  For instance, for $
t\rightarrow W^{+}b $ and $ \bar{t}\rightarrow W^{-}\bar{b} $ a
specific discrete symmetry implies a definite symmetry relation
among the associated helicity amplitudes.  In the case of $
\tilde{T}_{FS}$ invariance, the respective helicity amplitudes
must be purely real,
\begin{equation}
A^{*}\left( \lambda _{W^+},\lambda _b \right) =A\left( \lambda
_{W^+},\lambda _b \right)
\end{equation}
\begin{equation}
B^{*}\left( \lambda _{W^-},\lambda _{\bar b }\right) =B\left(
\lambda _{W^-},\lambda _{\bar b }\right)
\end{equation}
Intrinsic and relative signs of the helicity amplitudes are
specified in accordance with the standard Jacob-Wick phase
convention. $ \tilde{T}_{FS}$ invariance will be violated if
either (i) there is a fundamental violation of canonical ``time
reversal" invariance, or (ii) there are absorptive final-state
interactions. In the SM, there are no such final-state
interactions at the level of sensitivities considered in the
present analysis.  To keep this assumption of ``the absence of
final-state interactions" manifest, we refer to this as $
\tilde{T}_{FS}$ invariance, see [15] and references therein. If
experimental evidence for $ \tilde{T}_{FS}$ violation were found,
it would be very important to establish whether (i), (ii), or some
combination of the two effects was occurring. For instance,
unexpected final-state interactions might be associated with
addition t-quark decay modes.

To assess future measurements of helicity parameters in regard to
$ \tilde{T}_{FS}$ violation,  Figs. 1-5, are for the case of a
single additional pure-imaginary coupling, $ i g_i / 2 \Lambda_i $
or $ i g_i $, associated with a specific additional Lorentz
structure, $ i = S, P, S + P , \ldots $. In the SM, all the
relative phases are either zero or $\pm \pi$ so all of the primed
helicity parameters are zero.

\subsection{Additional $S \pm P, , f_M \pm f_E , S, P, f_M,$ or $f_E$ couplings}

The two plots displayed in Fig.1 are for dimensional couplings
with chiral $ S \pm P, f_M \pm f_E $ and non-chiral $ S, P, f_M,
f_E $ Lorentz structures. The upper plot displays the $
{\eta_L}^{'} $ helicity parameter versus the effective-mass scale
$ \Lambda_i $ with $g_i = 1 $ in $g_L = 1 $ units.  This parameter
is defined by
\begin{equation} \eta _L^{\prime }\equiv \frac 1\Gamma
|A(-1,-\frac 12)||A(0,- \frac 12)|\sin \beta _L
\end{equation}
where $ \beta_L=\phi _{-1}^L- \phi _0^L$ is the relative phase
difference the two helicity amplitudes in (5).  The peaks of the
curves shown in the upper plot do not however correspond to where
$\vert \sin \beta_L \vert \sim 1$.  Instead, at the peaks $\vert
\sin \beta _L \vert \sim 0.6-0.8$. The lower plot displays the
induced effect of the additional coupling on the partial width for
$ t\rightarrow W^{+}b $.  The SM model limit is at the ``wings"
where $ \vert \Lambda_i \vert \rightarrow \infty $ for each
additional dimensional coupling.  If the R-handed b-quark
amplitudes were found not to be negligible, it would be important
to consider both $ \eta _{L,R}^{\prime } $ or equivalently both $
\eta^{\prime }$ and $\omega^{\prime }$.

Fig.2 displays plots of the b-polarimetry interference parameters
${\epsilon_+}^{'}$ and $ {\kappa_0}^{'}$ versus $ \Lambda_i $ for
the case of a single additional  $ S, P, f_M, f_E $ and $ S \pm P,
f_M - f_E $ coupling.  These helicity parameters are defined by
\begin{equation}
\begin{array}{c}
\kappa _0^{\prime } \equiv \frac 1\Gamma |A(0,\frac
12)||A(0,-\frac 12)|\sin \alpha _0
\\
\epsilon _{+}^{\prime } \equiv \frac 1\Gamma |A(1,\frac 12)||A(0,-
\frac 12)|\sin \gamma _{+}
\end{array}
\end{equation}
where $ \alpha _0=\phi _0^R-\phi _0^L$ and $ \gamma _{+}=\phi
_1^R-\phi _0^L $ are the relative phases between the two
amplitudes in (6).  In the SM, the analogous $\kappa _0,\epsilon
_{+}$ in which the cosine function replaces the sine function are
the two ${\cal O}(LR)$ helicity parameters between the amplitudes
with the largest moduli.  Unfortunately, the tree-level values of
$\kappa _0,\epsilon _{+}$ in the SM are only about $1\%$.  Two
dimensional plots of the type $(\epsilon _{+},\eta _L)$ and
$(\kappa _0,\eta _L)$, and of their primed counterparts, have the
useful property that the unitarity limit is a circle of radius
$0.5$ centered on the origin.

In both plots, the peaks in the curves do correspond respectively
to where $\vert \sin \alpha_0 \vert   \sim 1$ and $\vert \sin
\gamma_{+} \vert \sim 1$, except that for $S,P$ in the lower $
{\kappa_0}^{'}$ plot where $\vert \sin \alpha_0 \vert   \sim 0.8$
at the peak.  The drops in the curves for small $ \vert \Lambda_i
\vert$'s is due to the vanishing of the sine of the corresponding
relative phase. Curves are omitted in the plots in this paper when
the couplings produce approximately zero deviations in the
helicity parameter of interest, e.g. this occurs for $ f_M + f_E $
in both the ${\epsilon_+}^{'}$ and $ {\kappa_0}^{'}$ helicity
parameters.

\subsection{Additional $V+A, V,$ or $A$ couplings}

An additional $V-A$ type coupling with a complex phase versus the
SM's $g_L$ is equivalent to an additional overall complex factor
in the SM's helicity amplitudes. This will effect the overall
partial width $\Gamma$, but it doesn't effect the other helicity
parameters.

For a single additional gauge-type coupling $V, A, $ or $V+A$, in
Fig.3 are plots of the b-polarimetry interference parameters
${\epsilon_+}^{'}$ and $ {\kappa_0}^{'}$, and of the partial width
for $ t\rightarrow W^{+}b $ versus pure-imaginary coupling
constant $ i g_i $.  The $g_i $ value is in $g_L = 1 $ units.  In
the cases of the additional dimensionless, gauge-type couplings,
the SM model limit is at the origin, $g_i \rightarrow 0 $. The
peaks for the $V+A$ coupling do correspond to where the associated
sine of the relative phase has maximum magnitude; instead, for the
$V, A$ couplings, $\vert \sin \alpha_0 \vert \sim 0.8$ $\vert \sin
\gamma_+ \vert \sim 0.8$ at the peaks.

\subsection{Indirect effects of $ \tilde{T}_{FS}$ violation on other helicity parameters}

The plots in Fig.4 show the indirect effects of a single
additional pure-imaginary chiral coupling, $ i g_i / 2 \Lambda_i $
or $ i g_i $, on other helicity parameters.  For the coupling
strength ranges listed in the ``middle table", the upper plot
shows the effects on the probability, $P(W_L)$, that the emitted $
W^{+} $ is ``Longitudinally" polarized and the effects on the
probability, $P(b_L)$, that the emitted b-quark has ``Left-handed"
helicity. Each curve is parametrized by the magnitude of the
associated $ g_i$ or $ \Lambda_i$.  On each curve, the central
open circle corresponds to the region with a maximum direct $
\tilde{T}_{FS}$ violation signature, e.g. for $ f_M + f_E $ from
Fig.1 this is at $ \vert \Lambda_{f_M + f_E} \vert \sim 50 GeV $.
The dark/light solid circles correspond respectively to the ends
of the ranges listed in the middle table where the direct
signatures fall to about $ 50\%$ of their maximum values.
Similarly the lower plot is for the W-polarimetry interference
parameters $ \eta, \omega$. Curves are omitted for the remaining
moduli parameter $ \zeta$, the pre-SSB parameter which
characterizes the odd-odd mixture of the $b$ and $W^+$
polarizations [15], because a single additional pure-imaginary
coupling in these ranges produces approximately zero deviations
from the pure $V-A$ value of $\zeta = 0.41 $.

The plots in Fig.5 show the indirect effects of a single
additional pure-imaginary non-chiral coupling on other helicity
parameters. Versus the middle table given here, the curves are
labeled as in Fig.4.

It is instructive to compare the above plots with their analogues
in [1].  Unlike in the analogous plots in [1], finite $m_b$
effects do not lead to sizable ``oval shapes" in plots in this
paper because interference terms must vanish in intensities
arising from the sum of a real $V-A$ amplitude and the
pure-imaginary $ i g_i / 2 \Lambda_i $ or $ i g_i $ amplitude.

In summary, sizable $ \tilde{T}_{FS} $ violation signatures can
occur for low-effective mass scales ( $ < 320 GeV $) as a
consequence of pure-imaginary couplings associated with a specific
additional Lorentz structure.  However, in most cases, such
additional couplings can be more simply excluded by $ 10\% $
precision measurement of the probabilities $ P(W_L)$ and $
P(b_L)$.  In most cases, the W-polarimetry interference parameters
$\eta$ and $\omega$ can also be used as indirect tests, or to
exclude such additional couplings.

\section{Tests for $ \tilde{T}_{FS}$ Violation Associated with the Dynamical Phase-Type Ambiguities}

The purpose of this section is to consider the situation when the
$ \tilde{T}_{FS}$ violation exists in the decay helicity
amplitudes, but nevertheless does not significantly show up in the
values of the moduli parameters.

Based on the notion of a complex effective mass scale parameter  $
\Lambda_{X} = \vert \Lambda_{X} \vert \exp{(-i \theta)} $ where
$\theta$ varies with the mass scale $\vert \Lambda_{X} \vert$, we
exploit the dynamical phase-type ambiguities to construct two
simple phenomenological models in which this happens. When $\sin
\theta $ $\geq 0$, the imaginary part of $ \Lambda_{X }$ could be
interpreted as crudely describing a more detailed/realistic
dynamics with a mean lifetime scale $\Gamma_{X} \sim 2 \vert
\Lambda_{X} \vert \sin \theta$ of pair-produced particles at a
production threshold $ Re [ 2  \Lambda_{X } ]$.  For $\sin \theta
$ $\leq 0$, fundamental time-reversal violation or a new dynamics
might approximately correspond to such a complex $ \Lambda_{X}$.
In the case of the $f_M+f_E$ ambiguity, over the full $\theta$
range, this construction does preserve the magnitudes' puzzle, see
Sec.1, between the $V-A$ and $f_M + f_E$ lines in the lower part
of Table 1.

{\bf $S+P$ dynamical, phase-type ambiguity:}

In Fig.6 are plots of the signatures for a partially-hidden $
\tilde{T}_{FS}$ violation associated with a $S+P$ phase-type
ambiguity. We require
$|A_{X}(0,-\frac{1}{2})|=|A_{L}(0,-\frac{1}{2})|$ to hold when the
additional $S+P$ coupling, $g_{S+P} / 2 \Lambda_{S+P} $ has a
complex effective mass scale parameter $ \Lambda_{S+P} = \vert
\Lambda_{S+P} \vert \exp{(-i \theta)} $ where $\theta$ varies with
the mass scale $\vert \Lambda_{S+P} \vert$. For $m_b = 0$, the
resulting relationship is $\cos \theta \simeq -\frac{%
m_{t}}{4\Lambda }(1-(\frac{m_{W}}{m_{t}})^{2})$ for $34.5GeV\leq
|\Lambda
_{S+P}|\leq \infty $ which correspond respectively to $\pm \pi \geq \theta \geq \pm \frac{%
\pi }{2}$. This construction maintains the standard model values
in the massless b-quark limit for the four moduli parameters, $
P(W_L), P(b_L) , \zeta,$ and $\Gamma$. The function $\theta (
\vert \Lambda_{S+P} \vert ) $ is then used for the $S+P$ coupling
when $ m_b = 4.5 GeV $. The SM values for the moduli parameters
are essentially unchanged. The phase choice of ${\phi^R}_1 = \pm
\pi$, cf. top line in Table 1, has no consequence since it is a $
2 \pi$ phase difference.

For $ \sin{ \theta} \geq 0 $, in Fig. 6 the solid curve shows $
{\eta_L}^{'} $ plotted versus $ 1 / \vert \Lambda_{S+P} \vert$.
The dashed curve is for $\eta_L, \eta, \omega$ which are
degenerate. The dark rectangles show the standard model values at
the $\vert \Lambda_{S+P} \vert\rightarrow \infty $ endpoint.  At
the other endpoint $\vert \Lambda_{S+P} \vert \sim 34.5 GeV $, or
$ 1 / \vert \Lambda_{S+P} \vert = 0.029 GeV^{-1}$.

From the perspectives of (i) measuring the $W$ interference
parameters and of (ii) excluding this type of $ \tilde{T}_{FS}$
violation, it is noteworthy that where $ {\eta_L}^{'} $ has the
maximum deviation, there is a zero in $\eta_L, \eta, \omega$.  So
if the latter parameters were found to be smaller than expected or
with the opposite sign than expected, this would be consistent
with this type of $ \tilde{T}_{FS}$ violation.

At the maximum of $ {\eta_L}^{'} $, $ \vert \Lambda_{S+P} \vert
\sim 49 GeV$ and the other $ \tilde{T}_{FS}$ violation parameters
are also maximum. The curves for these parameters have the same
over all shape as  $ {\eta_L}^{'} $ but their maxima are small,
${\epsilon_+}^{'} \sim 0.015$ and ${\kappa_0}^{'} \sim 0.028$. For
the other case where $ \sin{\theta } \leq 0 $, all these $
\tilde{T}_{FS}$ violation primed parameters have the opposite
overall sign.  The signs of other helicity parameters are not
changed.

{\bf $f_M+f_E$ dynamical, phase-type ambiguity:}

In Fig. 7 are plots of the signatures for a partially-hidden $
\tilde{T}_{FS}$ violation associated with a $f_M+f_E$ phase-type
ambiguity. As above for the analogous $S+P$ construction, the
additional $f_M+f_E$ coupling $g_{f_M+f_E} / 2 \Lambda_{f_M+f_E} $
now has an effective mass scale parameter $ \Lambda_{f_M+f_E} =
\vert \Lambda_{f_M+f_E} \vert \exp{(-i \theta)} $ in which
$\theta$ varies with the mass scale $\vert \Lambda_{f_M+f_E}
\vert$ to maintain standard model values in the massless b-quark
limit for
the moduli parameters $ P(W_L), P(b_L) ,$ and $ \zeta$.
For $X=f_{M}+f_{E}$, we require $\frac{|A_{X}(-1,-\frac{1}{2})|}{|A_{X}(0,-%
\frac{1}{2})|}=\frac{|A_{L}(-1,-\frac{1}{2})|}{|A_{L}(0,-\frac{1}{2})|}$
so for $m_{b}=0$ the relationship giving $\theta ( \vert \Lambda_{
f_{M}+f_{E} } \vert ) $  is $\cos \theta \simeq
\frac{m_{t}}{4\Lambda }(1+(\frac{m_{W}}{m_{t}})^{2})$ for
$52.9GeV\leq |\Lambda _{f_{M}+f_{E}}|\leq \infty $ which
correspond respectively to
 $ 0\leq \theta \leq \pm \frac{\pi }{2}$.  For the case $ \sin{
\theta} \geq 0 $,  in Fig.8 the upper plot shows by the solid
curve $ {\eta_L}^{'} $ versus $ 1 / \vert \Lambda_{f_M+f_E}
\vert$.  By the dashed curve, it shows  $\eta_L, \eta, \omega$. At
the endpoint $\vert \Lambda_{f_M+f_E} \vert \sim 52.9 GeV $, or $
1 / \vert \Lambda_{f_M+f_E} \vert = 0.0189 GeV^{-1}$.

Here, as in Fig. 7, where $ {\eta_L}^{'} $ has the maximum
deviation, there is a zero in $\eta_L, \eta, \omega$. The lower
plot shows the indirect effect of such a coupling on the partial
width $\Gamma$.  While $ \vert \Lambda_{f_M+f_E} \vert$ varies,
two of the relative phases remain almost fixed, $\gamma_{\pm} \sim
\pm \pi $ respectively, so only one independent relative phase
could be viewed as driving the variation, e.g. $\beta_L$ varies
from $ - \pi$ to zero.

At the maximum of $ {\eta_L}^{'} $, $ \vert \Lambda_{f_M+f_E}
\vert \sim 63 GeV$.  The curve for ${\kappa_0}^{'}$ has the same
shape and is also maxmimum at the same position with a value
${\kappa_0}^{'} \sim 0.005$. ${\epsilon_+}^{'} $ remains very
small.  For the other case, $ \sin{\theta } \leq 0 $, each of
these $ \tilde{T}_{FS}$ violation primed parameters has the
opposite overall sign.  Since sign$(-\sin \phi _{0}^{L})=$
sign$(\sin \phi _{0}^{R})=$sign$(\sin \theta )$ and sign$(-\sin
\phi _{-1}^{L})=$ sign$(\sin \phi _{1}^{R})=$sign$(\sin \theta )$,
all the relative phases change sign for the case $ \sin{\theta }
\leq 0 $.

In summary, sufficiently precise measurement of the W-interference
parameters $\eta_L$ and $ {\eta_L}^{'} $ can exclude such
partially-hidden $ \tilde{T}_{FS}$ violation associated with
either of the two dynamical phase-type ambiguities. However, if
$\eta_L = ( \eta + \omega ) / 2 $ were found to be smaller than
expected or with a negative sign, such a measurement would be
consistent with this type of $ \tilde{T}_{FS}$ violation.

\begin{center}
{\bf Acknowledgments}
\end{center}

For computer assistance, one of us (CAN) thanks John Hagan and Ted
Brewster. This work was partially supported by U.S. Dept. of
Energy Contract No. DE-FG 02-86ER40291.

\newpage

\begin{center}
{\bf Table Captions}
\end{center}

Table 1: For the standard model and at the ambiguous moduli
points, numerical values of the associated helicity amplitudes $
A\left( \lambda_{W^{+} } ,\lambda_b \right) $. The values for the
amplitudes are listed first in $ g_L = 1 $ units, and second as $
A_{new} = A_{g_L = 1} / \surd \Gamma $ which removes the effect of
the differing partial width,
$
\Gamma $ for $ t\rightarrow W^{+}b $. [$m_t=175GeV, \;
m_W =
80.35GeV, \; m_b = 4.5GeV$ ].

\begin{center}
{\bf Figure Captions}
\end{center}

FIG. 1: The first {\bf five sets of figures} are for the case of a
single additional pure-imaginary coupling, $ i g_i / 2 \Lambda_i $
or $ i g_i $, associated with a specific additional Lorentz
structure, $ i = S, P, S + P , \ldots $.  The two plots displayed
here are for dimensional couplings with chiral $ S \pm P, f_M \pm
f_E $ and non-chiral $ S, P, f_M, f_E $ Lorentz structures. The
{\bf upper plot} displays the $ {\eta_L}^{'} $ helicity parameter
versus the effective-mass scale $ \Lambda_i $ with $g_i = 1 $ in
$g_L = 1 $ units. The {\bf lower plot} displays the induced effect
of the additional coupling on the partial width for $ t\rightarrow
W^{+}b $.  The standard model(SM) limit is at the ``wings" where $
\vert \Lambda_i  \vert \rightarrow \infty $.

FIG. 2: Plots of the b-polarimetry interference parameters
${\epsilon_+}^{'}$ and $ {\kappa_0}^{'}$ versus $  \Lambda_i $ for
the case of a single additional  $ S, P, f_M, f_E $ and $ S \pm P,
f_M - f_E $ coupling.  Curves are omitted in the plots in this
paper when the couplings produce approximately zero deviations in
the helicity parameter of interest.

FIG. 3: For a single additional gauge-type coupling $V, A, $ or
$V+A$, plots of the b-polarimetry interference parameters
${\epsilon_+}^{'}$ and $ {\kappa_0}^{'}$, and of the partial width
for $ t\rightarrow W^{+}b $ versus pure-imaginary coupling
constant $ i g_i $.  The  $g_i $ value is in $g_L = 1 $ units. The
SM model limit is at the origin, $g_i \rightarrow 0 $.

FIG. 4:  These plots show the indirect effects of a single
additional pure-imaginary chiral coupling, $ i g_i / 2 \Lambda_i $
or $ i g_i $, on other helicity parameters.  A dark rectangle
denotes the value for the SM.  For the coupling strength ranges
listed in the ``middle table", the {\bf upper plot} shows the
indirect effects on the probabilities $P(W_L)$ and $P(b_L)$. Each
curve is parametrized by the magnitude of the associated $ g_i$ or
$ \Lambda_i$.  On each curve, the central open circle corresponds
to the region with a maximum direct $ \tilde{T}_{FS}$ violation
signature. The dark/light solid circles correspond respectively to
the ends of the ranges listed in the middle table where the direct
signatures fall to about $ 50\%$ of their maximum values. The {\bf
lower plot} is for the W-polarimetry interference parameters $
\eta, \omega$.

FIG. 5:  These plots show the indirect effects of a single
additional pure-imaginary non-chiral coupling on other helicity
parameters.  Versus the middle table given here, the curves are
labeled as in Fig. 4. The {\bf upper plot} is for the two
probabilities $P(W_L)$ and $P(b_L)$. The {\bf lower plot} is for $
\eta, \omega$.

FIG. 6:  Plots of the signatures for a partially-hidden $
\tilde{T}_{FS}$ violation (see text) associated with a $S+P$
phase-type ambiguity.  Plotted versus $ 1 / \vert \Lambda_{S+P}
\vert$ for the case $\sin {\theta} \geq 0$  is the solid curve for
$ {\eta_L}^{'} $, and the dashed curve for $\eta_L, \eta, \omega$
which are degenerate.

FIG. 7:  Plots of the signatures for a partially-hidden $
\tilde{T}_{FS}$ violation (see text) associated with a $f_M+f_E$
phase-type ambiguity. Versus $ 1 / \vert \Lambda_{f_M+f_E} \vert$
for $\sin {\theta} \geq 0$, the {\bf upper plot} shows by the
solid curve $ {\eta_L}^{'} $, and by the dashed curve $\eta_L,
\eta, \omega$.  The {\bf lower plot} shows the indirect effect of
such a coupling on the partial width.


\begin{thebibliography}{333}
\bibitem{1} C.A. Nelson and A.M. Cohen, Eur. Phys. J. {\bf C8},
393(1999).   In the case of the $S+P$ ambiguity, Table 2 should
list $\kappa_0 = 0.05 $, i.e. with a positive sign.
\bibitem{2} The CDF collaboration reported that the fraction of
longitudinal W bosons is $F_0 = 0.91 \pm 0.37 \pm 0.13$ assuming a
pure $V-A$ coupling, T. Affolder, et.al., Phys.Rev.Lett. {\bf 84},
216(2000); the D{\O} collaboration reported $t \bar{t}$
spin-correlation results in S. Snyder, FERMILAB-CONF-99-294-E,
hep-ex/9910029, to appear in Proc. EPS-HEP 99.
\bibitem{3} M.Beneke, {\it et. al.}, hep-ph/0003033, to appear in ``1999 CERN Workshop on SM physics (and more) at the LHC".
\bibitem{4} D. Atwood, S. Bar-Shalom, G. Eilam, and A. Soni, hep-ph/0006032.
\bibitem{5} Reports from a top quark workshop at Fermilab include G. Mahlon hep-ph/9811219,
hep-ph/9811281; S.S. Willenbrock, hep-ph/9905498.
\bibitem{6} W. Bernreuther, A. Brandenburg, and Z.G. Si, Phys.Lett. {\bf B483},
99(2000);  W. Bernreuther, A. Brandenburg, and M. Flesch,
hep-ph/9812387.
\bibitem{7}  M. Fischer, S. Groote, J.G. K\"{o}rner, M.C. Mauser, B.
Lampe, Phys.Lett. {\bf B451}, 406(1999);  B. Lampe,
hep-ph/9801346.
\bibitem{8} B. Grzadkowski, Z. Hioki, Phys.Rev. {\bf D61}, 014013(2000); hep-ph/0003294.
\bibitem{9} G. Mahlon and S. Parke, Phys.Lett. {\bf B476}, 323(2000);
hep-ph/0001201; J. Kodaira, T. Nasuno, S. Parke,  hep-ph/9807209.
\bibitem{10} T. Tait, Phys. Rev. {\bf D61}, 0340001(2000); F. Larios and C.-P. Yuan, Phys.Lett. {\bf B457}, 334(1999);  T. Tait
and C.-P. Yuan, hep-ph/9710372.
\bibitem{11} Y. Kiyo, J. Kodaira, K. Morii, T. Nasuno, and S. Parke, hep-ph/0006021.
\bibitem{12} J. Guasch, W. Hollik, J. I. Illana, C. Schappacher, and J. Sola,
hep-ph/0003109.
\bibitem{13} E. Boos, M. Dubinin, M. Sachwitz, and H.J. Schreiber,
hep-ph/0001048; A.Belyaev and E. Boos, hep-ph/0003260; A.Belyaev,
hep-ph/0007058.
\bibitem{14} M. Jack, A. Hoefer, A. Leike, and T. Riemann,
hep-ph/0007046.
\bibitem{15} C.A. Nelson, B.T. Kress, M. Lopes, and T.P. McCauley, Phys. Rev.
{\bf D56}, 5928(1997); {\bf D57}, 5923(1998).

\end{thebibliography}
\end{document}